\documentstyle[aps,prl]{revtex}

\title{Good dynamics versus bad kinematics. Is entanglement needed for 
quantum 
computation?}

\author{Noah Linden$^{a}$ and
Sandu Popescu$^{a,b}$}

\address{
$^a$Isaac Newton Institute for Mathematical
Sciences, 20 Clarkson Road, Cambridge, CB3 0EH, UK.\\
$^b$BRIMS, Hewlett-Packard Laboratories, Stoke Gifford, 
Bristol BS12 6QZ, UK}

\date{3 June 1999}

\begin{document}

\twocolumn

\draft

\maketitle

\begin{abstract}
We consider quantum computing with pseudo-pure states.  This framework
arises in certain implementations of quantum computing using NMR.
We analyze quantum computational protocols which aim to solve exponential 
classical problems with polynomial resources and ask whether or not 
entanglement of the pseudo-pure states is needed to achieve this aim. We 
show that for a large class of such protocols, including Shor's 
factorization, entanglement is necessary.  
We also show that achieving entanglement is not sufficient: if the noise 
in the state is sufficiently large, exponential resources are needed even 
if entanglement is present.

\end{abstract}


In a beautiful example of how technology can stimulate fundamental 
physics, 
the proposals for implementing quantum computing via liquid state NMR 
\cite{Cory97a,Cory97b,Gershenfeld97,Chuang97} have 
 sparked a debate recently on  the very nature of quantum 
computing\cite{Braunstein99,Laflamme,Cory98,Caves}. 
More precisely, doubts have been raised as to whether entanglement is a 
necessary 
requirement for a quantum computer to be able to  
speed-up 
a computation (exponentially) relative to a classical computer\cite{Laflamme}. 

The proposal to use liquid state 
NMR with pseudo-pure states 
for quantum computing has two important aspects. 

\begin{itemize}
\item
{\bf Bad kinematics.} On the one hand liquid state NMR quantum computing 
has 
a great disadvantage: One 
cannot 
prepare 
pure states. 
This situation is different from the original quantum computation 
protocols 
which considered the quantum computer in a pure state $|\Psi\rangle$.
Instead, in the NMR protocol one prepares ``pseudo-pure" states, i.e. 
mixed 
states of the form

\begin{equation}
\rho=(1-\epsilon)M+\epsilon|\Psi\rangle\langle\Psi| \label{mixedrho}
\end{equation}
where $M$ is the maximally mixed state (i.e. the identity density matrix 
normalized to have trace 1). In other words, in NMR the pure state 
$|\Psi\rangle$ 
is 
contaminated by noise.

Furthermore  it has been shown \cite{Braunstein99} that in 
all 
experiments to date the noise is so large that the ``pseudo-pure" state 
$\rho$ 
is {\em non-entangled} even if the pure-state component $|\Psi\rangle$ is 
entangled.
Since entanglement is widely considered to be the main ingredient in 
quantum 
computing, these results lead to the question as to whether the NMR 
scheme is  a ``true" 
quantum 
computation \cite{Braunstein99}.

\item 
{\bf Good dynamics} On the other hand, the NMR experiments have a great 
advantage: one can 
produce 
{\it correct dynamics}, that is, the interactions between the spins are 
exactly 
as required in the theoretical quantum computational protocols. Thus if 
the 
initial state of the spins would be pure instead of ``pseudo-pure", NMR 
experiments would completely implement the original quantum computation 
protocols.

Furthermore, the noise in the ``pseudo-pure" state looks quite benign - 
it 
averages to zero (without loss of generality we
can consider our observables to be traceless). Thus the expectation value of any operator $A$ when the 
quantum 
system is in a given pure state $\Psi$ is the same, up to normalization, 
as 
the 
average in the corresponding  pseudo-pure state 
$\rho=(1-\epsilon)M+\epsilon|\Psi\rangle\langle\Psi|$, i.e.

\begin{equation}
Tr (A\rho)=\epsilon\langle\Psi|A|\Psi\rangle
\end{equation}

\end{itemize}
Given the good dynamics some authors have suggested that in fact 
liquid-state 
NMR 
computing is nonetheless a ``true" quantum computation, capable of 
speeding-up 
computations relative to classical computers. As a corollary, it was 
suggested 
that perhaps entanglement is {\em not} a {\it sine-qua-non} requirement 
for 
quantum 
computing \cite{Laflamme}.

Whether or not entanglement is a necessary condition for quantum 
computation is 
a question of fundamental importance, which obviously goes far beyond the 
NMR 
computing context in which it arose. In the present Letter we study this 
question for the pseudo-pure state quantum computing.
Specifically we analyze quantum computational protocols which aim to solve 
exponential classical problems with polynomial resources and ask whether 
or not entanglement of the pseudo-pure states is needed to achieve this 
aim. We show that for a large class of such protocols, including Shor's 
factorization\cite{Shor}, entanglement of the pseudo-pure states is necessary:  
 unless 
the pseudo-pure state (\ref{mixedrho}) of the quantum computer becomes 
entangled during the 
computation, the aim of transforming exponential problems to polynomial 
ones cannot be achieved. 

We will first consider the general effect of noise 
on the computation. Then the  relationship between 
separability and noise.

Consider then a pure state computational protocol
in which the computer starts in the state $|\Psi_0\rangle$
and ends in the state $|\Psi_{f}\rangle = U|\Psi_0\rangle$
where $U$ is the unitary time evolution operator which describes
the computation.  
The corresponding computation starting with pseudo-pure
state 
\begin{equation}
\rho=(1-\epsilon)M+\epsilon|\Psi_0\rangle\langle\Psi_0| 
\end{equation}
ends up in the state
\begin{equation}
\rho=(1-\epsilon)M+\epsilon|\Psi_f\rangle\langle\Psi_f| 
.\label{finalmixed}
\end{equation}
Upon reaching the final state, a measurement is carried out
and the result of the computation is inferred from the result
of the measurement.

We will assume the most favorable case that the pure state protocol gives 
the correct
answer with certainty with a single repetition of the 
protocol and that if the result of the computation is 
found, 
one can check it with polynomial overhead.  
We will then show that the pseudo-pure state
protocol requires of the order of $1\over \epsilon$ repetitions.
Thus if $\epsilon$ becomes exponentially small with $N$, the number 
governing the scaling of the classical problem, (in other
words the noise becomes exponentially large with $N$), the protocol
requires an exponential number of repetitions to get the correct answer.
So for this amount of noise, the quantum protocol with a pseudo-pure 
state cannot transform
a exponential problem into a polynomial one: 
even in the best possible case that  the pure-state protocol
takes one computational step, the protocol with
noise takes exponentially many steps. We emphasize that this conclusion applies  
quite 
generally to pseudo-pure state quantum computing  and is independent of the 
discussion of separability which follows later.

In the state (\ref{finalmixed}) there is a probability
$\epsilon$ of finding the computer in the ``correct'' final
state $|\Psi_f\rangle$ arising from the term
\begin{equation}
\epsilon|\Psi_f\rangle\langle\Psi_f|
\end{equation}
in (\ref{finalmixed}).  As stated above, will assume here the 
most favorable case, that if the state is $|\Psi_f\rangle$ 
then from the outcome of the final measurement, one 
can infer the solution to the computational problem with 
certainty with one repetition.  We note that general protocols, such as 
Shor's algorithm
for example,
a single repetition of the protocol is not sufficient to find 
the correct answer.

There is also  the probability 
$(1-\epsilon)$ of 
finding the computer in the maximally mixed state $M$.  In this case
there {\em is} a possibility that the correct answer will be found, since
the noise term contains all possible outcomes with some probability.  
However, the probability of finding the correct answer from the noise
term must be at least exponentially small with $N$.  Otherwise
there would be no need to prepare the computer at all:  one could
find the correct answer from the noise term simply by repeating the 
computation a polynomial number of times.  In fact, if the probability of 
finding the correct answer from the noise
term did not become exponentially small with $N$ we could dispense with 
the computer altogether.  For using a classical probabilistic protocol 
which selected from all the  possibilities at random, we would get the 
correct answer with probability of the order of one with only a polynomial 
number of trials.

Thus we may say that the probability of finding the correct answer
from the state (\ref{finalmixed}) is essentially $\epsilon$ and so
the computation must be repeated $1\over\epsilon$ times on average to 
find the correct answer with probability of order one.  

We now consider whether reaching entangled states during the computation 
is a {\em necessary} condition for exponential speed-up.  We address this 
by 
investigating what can be achieved with {\em separable} states.
Specifically  we impose the condition that the pseudo-pure state remains 
separable during the entire computation. For a important class of 
computational protocols we show that this condition implies an exponential 
amount of noise.

The protocols which we consider use  $n=n_1+n_2$ 
qubits
of which $n_1$ are considered to be the input registers and the remaining
$n_2$, the output registers.  We assume that  $n_1$ and $n_2$
are polynomial in the 
number $N$ which describes how the classical problem scales.    
As stated 
earlier we consider problems in which the quantum protocol gives and 
exponential speed-up over the classical protocol, specifically the 
classical protocol is exponential in $N$ whereas the quantum protocol is 
polynomial in $N$. (For example in the factorization problem, the aim is 
to factor a number of the order of $2^N$.  The classical protocol is 
exponential in $N$ and in Shor's algorithm, $n_1$ and $n_2$ are 
 linear in $N$.)

We first describe the protocols as applied to pure states. The first steps 
are
\begin{itemize}
\item Prepare system in the ground state
\begin{eqnarray}
|\Psi_0\rangle = |00....0\rangle\otimes | 00..0\rangle\label{Psi0}
\end{eqnarray}
\item Perform a Hadamard transform on the input register,
so that the state becomes 
\begin{eqnarray}
|\Psi_1\rangle = {1\over 2^{{n_1}/2}} \sum_{x=0}^{2^{n_1}-1} |x\rangle 
\otimes | 00..0\rangle
\end{eqnarray}
\item Evaluate the function $f:\ \{0,1\}^{n_1} \rightarrow \{0,1\}^{n_2}$. The 
state becomes
\begin{eqnarray}
|\Psi_2\rangle = {1\over 2^{n_1/2}} \sum_{x=0}^{2^{n_1}-1} |x\rangle 
\otimes 
| f(x)\rangle
\end{eqnarray}
\end{itemize}
Now consider the protocol when applied to a mixed state input.
Thus the initial state $\rho_0$ is
\begin{eqnarray}
\rho_0 = (1-\epsilon){ M_{2^{n}}} + \epsilon 
|\Psi_0\rangle 
\langle \Psi_0| \label{mixedPsi0}
\end{eqnarray}
where $|\Psi_0\rangle $ is given in (\ref{Psi0}), and
$M_{2^{n}}$ is the maximally mixed state in the $2^n$ dimensional
Hilbert space.
After the second computational step the state is
\begin{eqnarray}
\rho_0 = (1-\epsilon){ M_{2^{n}}} + \epsilon 
|\Psi_2\rangle 
\langle \Psi_2|.\label{mixedPsi2}
\end{eqnarray}
Consider now protocols in which the function $f(x)$ is not constant. Let 
$x_1$ and $x_2$ be values of $x$ such that $f(x_1) \neq f(x_2)$.  Thus we 
may write the state $|\Psi_2\rangle$ as
\begin{eqnarray}
|\Psi_2\rangle = {1\over 2^{n_1/2}} \left( |x_1\rangle |f(x_1)\rangle + 
|x_2\rangle 
|f(x_2)\rangle + |\Psi_r\rangle \right)
\end{eqnarray}
where $|\Psi_r\rangle$ has no components in the subspace spanned by
$|x_1\rangle |f(x_1)\rangle,\ |x_1\rangle |f(x_2)\rangle,\ |x_2\rangle 
|f(x_1)\rangle,\ 
|x_2\rangle |f(x_2)\rangle$.  It is convenient to relabel
these states and write
\begin{eqnarray}
|\Psi_2\rangle = {1\over 2^{n_1/2}} \left( |1\rangle |1\rangle + |2\rangle 
|2\rangle + |\Psi_r\rangle \right)
\end{eqnarray}
where $|\Psi_r\rangle$ has no components in the subspace spanned by
$|1\rangle |1\rangle,\ |1\rangle |2\rangle,\ |2\rangle |1\rangle,\ 
|2\rangle |2\rangle$.

We now derive a necessary condition on $\epsilon$ for the state of the 
system to be separable throughout the computation.  For consider
projecting each particle onto the subspace spanned by 
$|1\rangle$ and 
$|2\rangle$.  The state after projection is
\begin{eqnarray}
\rho_2^\prime&=& 
{1\over A}
\biggl[{4(1-\epsilon)\over 2^{n_1+n_2}}{M_{4}} \nonumber\\
& &\qquad+{2\epsilon\over 2^{n_1}}
\Bigl(
{|1\rangle|1\rangle+|2\rangle|2\rangle \over \sqrt 2}
\Bigr)
\Bigl(
{\langle 1|\langle 1| + \langle 2|\langle 2|\over \sqrt 2}
\Bigr)
\biggr]
\nonumber\\
&=&  
(1-\epsilon'){M_{4}}\nonumber\\
& & \quad +\epsilon'\Bigl(
{|1\rangle|1\rangle+|2\rangle|2\rangle \over \sqrt 2}
\Bigr)
\Bigl(
{\langle 1|\langle 1| + \langle 2|\langle 2|\over \sqrt 2}
\Bigr)
\;,
\label{entangled}
\end{eqnarray}
where 
\begin{eqnarray}
A=\left( {4(1-\epsilon)\over 2^{n_1+n_2}}+ {2\epsilon\over 
2^{n_1}}\right)
\end{eqnarray}
 is the normalization factor,  $M_{4}$ is the maximally mixed state in the 
4-dimensional Hilbert space spanned by $|1\rangle |1\rangle,\ |1\rangle 
|2\rangle,\ |2\rangle |1\rangle,\ 
|2\rangle |2\rangle$
 and 
\begin{equation}
\epsilon'=
{2\epsilon\over 2^{n_1} A}=
{\epsilon \over (1-\epsilon)2^{-n_2+1} + \epsilon }
.
\end{equation}
Now a 2 qubit state of the form
\begin{equation}
(1-\delta){M_{4}}+\delta\Bigl(
{|1\rangle|1\rangle+|2\rangle|2\rangle \over \sqrt 2}
\Bigr)
\Bigl(
{\langle 1|\langle 1| + \langle 2|\langle 2|\over \sqrt 2}
\Bigr)
\end{equation}
is entangled for $\delta > 1/3$.  Therefore the original state 
(\ref{mixedPsi2}) must have been entangled 
unless 
\begin{equation}
\epsilon^\prime \leq 1/3 \quad \Rightarrow \quad \epsilon \leq 
{1\over 1+2^{n_2}},
\end{equation}
since local projections cannot create entangled states from unentangled 
ones.

Therefore we conclude that if we have a computational protocol (for 
non-constant $f$) starting with  a mixed state of the form 
(\ref{mixedPsi0}) and if we require that the state remains separable 
throughout the protocol, then we certainly need
\begin{equation}
 \epsilon \leq { {1\over 1+2^{n_2}}}.
\end{equation}
However we have shown earlier that, even in favorable circumstances, a 
computation with noise $\epsilon$ takes of the order of $1/\epsilon$
repetitions to get the correct answer with probability of the order of 
one.

Thus we reach our main result that computational protocols of the sort we 
have considered require exponentially many repetitions. So no matter how 
efficient the original pure state protocol is, the mixed state protocol 
which is sufficiently noisy that it remains separable for all $N$, will 
{\em not} 
transform an exponential classical problem into a polynomial one.  

We note that
while we have considered protocols of a specific form, many of the details
are unimportant. As long as the number of qubits 
$n_1$ and $n_2$ in the
output register is 
polynomial in the number $N$ which governs the classical problem, and the 
pure state protocol goes through a state which has a non-negligible amount 
of entanglement, similar conclusions can be drawn.

We repeat here that our conclusions only apply to separable states of 
pseudo-pure state
form (\ref{mixedrho}). We have nothing to say at this
stage about separable states of other forms.
We have also only considered exponential speed up so that our results do 
not apply to Grover's algorithm for example \cite{Grover}. Furthermore, we 
cannot rule out the possibility of the future discovery of more efficient 
algorithms of Shor type to which our results do not apply.

We have shown earlier that having entanglement is a necessary condition: 
is it sufficient?  Our earlier results show that it is not.  As long as 
the noise is exponential (i.e. $\epsilon$ decreases exponentially with 
$n$), the computation has to be repeated an exponential number of times,
even if entangled states are reached during the computation (we note that it is 
known that there are entangled states with an 
exponential amount of noise\cite{Braunstein99}). 

Finally let us return to NMR quantum computation which gave rise to the 
issues we have been discussing. 
In our previous discussion we had in mind that one has a single quantum 
computer and $1/\epsilon$ then gives the number of times the computation 
has to be repeated. In NMR, 
each molecule in the sample is considered to be a quantum computer. Here,
rather than repeating the computation on the same computer, one treats
a large number of computers in parallel. Thus, if each molecule could be 
accessed individually, $1/\epsilon$ would the number of individual 
molecular computers one would need in the sample.  In fact, rather than being 
able to address each molecule 
individually,  one can only 
measure bulk properties of the sample  - a far less favorable situation -
so that $1/\epsilon$ is a lower bound on the sample scaling required.

The usual construction of pseudo-pure states in NMR 
has $\epsilon \sim { n\over 2^n}$.  In the light
of the discussion in the previous paragraph the sample size would 
therefore have to grow exponentially with $n$.  Thus this framework
does not allow one to convert  exponential classical problems into    
polynomial ones via Shor-type protocols.

Of course,  other ways of using liquid state NMR as a quantum computer 
might be found with more favorable scaling than current techniques.  
Whether or not this is possible is obviously beyond the scope of this 
paper.

\bigskip
\noindent{\large\bf Acknowledgments}

We are grateful to Sam Braunstein, Carlton Caves, Richard Jozsa and Ruediger Schack 
for many illuminating 
discussions and to David Collins for pointing out
an error in an earlier draft.


\begin{thebibliography}{99}
\bibitem{Cory97a} 
D.~G. Cory, A.~F. Fahmy, and T.~F. Havel,
Proc. Nat. Acad. Sci. USA {\bf 94}, 1634 (1997).
%
\bibitem{Cory97b} 
D.~G. Cory, M.~D. Price, and T.~F. Havel,
Physica D {\bf 120}, 82 (1998).
%
\bibitem{Gershenfeld97} 
N.~Gershenfeld  and I.~L. Chuang,
Science {\bf 275}, 350 (1997).
%
\bibitem{Chuang97} 
I.~L. Chuang,  N.~Gershenfeld, M.~G. Kubinec, and D.~W. Leung,
Proc. R. Soc. Lond. A {\bf 454}, 447 (1998).
%
\bibitem{Braunstein99} S.~L.~Braunstein, C.~M.~Caves,  R.~Jozsa,  
N.~Linden, 
S.~Popescu,  and R.~Schack, quant-ph/9811018.
%
\bibitem{Laflamme} R.~Laflamme, Review of ``Separability of very noisy mixed 
states and implications for NMR quantum\hfil\break computing" by S. Braunstein et al., in 
Quick reviews in quantum computation
and information,\hfil\break http://quantum-computing.lanl.gov/qcreviews/qc/. 
%
\bibitem{Cory98} T.~F.~Havel, S.~S.~Somaroo, C.-H.~Tseng, and D.~G.~Cory, 
quant-ph/9812086.
%
\bibitem{Caves}R.~Schack and C.~M.~Caves, quant-ph/9903101.
%
\bibitem{Shor} P.~Shor, in {\it Proceedings of 35th Annual Symposium on the 
Foundations of Computer Science}, ed. S. Goldwasser (IEEE Comp. Soc. Press, Los 
Alamitos, CA) p 116 (1994).
%
\bibitem{Grover} L. Grover, Phys. Rev. Lett {\bf 78}, 325 (1997).
%
\end{thebibliography}
\end{document}